\documentclass[12pt,a4paper]{article}   
\usepackage{amsmath,amsfonts,amssymb}
\usepackage{latexsym}
\newcommand{\np}{\nonumber\\}
\newtheorem{thm}{Theorem}
\newtheorem{lem}{Lemma}

\title{
Anharmonic Oscillators, Spectral Determinant and
Short Exact Sequence of $U_q(\widehat{\mathfrak{sl}}_2)$
}
\author{ J. Suzuki\thanks{e-mail: suz@hep1.c.u-tokyo.ac.jp}
                   \thanks{Address after April 1999:
                   Department of Physics, Faculty of Science,
                   Shizuoka University, 836 Ohya, Shizuoka 422, Japan.}\\
        \parbox{0.9\textwidth}{
        {\em
        \begin{center}
       Institute of Physics,\\
        University of Tokyo at Komaba,\\
       Komaba 3-8-1, Meguro-ku, Tokyo,\\
       Japan
        \end{center}
        }}
       }
\begin{document}
\maketitle
\begin{abstract}
We prove one of conjectures, raised by Dorey and Tateo 
in the connection among the 
spectral determinant of anharmonic oscillator and
vacuum eigenvalues of transfer matrices in field theory and
statistical mechanics.
The exact sequence of $U_q(\widehat{\mathfrak{sl}}_2)$ plays a 
fundamental role in the proof.
\end{abstract}
\noindent pacs number: 05.50.+q,11.55.Ds.
\clearpage

Recently, Dorey and Tateo 
 have found a remarkable connection among
the spectral determinants of a 1D Schr{\"o}dinger operator
associated with the anharmonic oscillator, 
transfer matrices and ${\bf Q}$ operators in CFT 
 for a certain value of
Virasoro parameter $p$\cite{DT}.
This has been subsequently generalized to general values of
$p$ by appropriate modifications on the Hamiltonian\cite{BLZnew}.
The most fundamental equalites among parity dependent spectral 
determinants and 
${\bf Q}_{\pm}$ operators are proven by utilizing 
the quantum Wronskian relation. 

In this note, we provide an elementary proof of the 
conjectures in \cite{DT} concerning the sum rule which is
closed only among the 
spectral determinant ($=$ product of
  parity dependent spectral determinants).
The short exact sequence in
quantum affine Lie algebra $U_q(\widehat{\mathfrak{sl}_2})$ 
plays a fundamental role.

We consider 
the Schr{\"o}dinger equation,

\begin{equation}
\widehat{H} \Psi_k(x)=
\Bigl( -\frac{d^2}{dx^2} + x^{2M} \Bigr ) \Psi_k(x)= E_k \Psi_k(x),
\end{equation}
Here $M$ is assumed to be an integer greater than 2.

The spectral problem associated with this has been scrutinized 
in \cite{V1}-\cite{V5}.
The properties can be encoded into the spectral determinant
\begin{equation}
D_M(E) = {\rm det}(E+\widehat{H})=
D_M(0) \prod_{k=0}^{\infty} \Bigl( 1+\frac{E}{E_k} \Bigr ),
\end{equation}
and $D_M(0)=1/\sin(\pi/(2M+2))$.

In the following, we adopt a notation
${\cal D}_M(x): =D_M( {\rm e}^{\pi x/(M+1)})$.

Remarkably, it satisfied the exact functional relation \cite{V2},
which reduces to a simple polynomial form for $M=2$
\begin{equation}
{\cal D}_2(x) {\cal D}_2(x+2i) {\cal D}_2(x+4i) = 
{\cal D}_2(x) + {\cal D}_2(x+2i)+{\cal D}_2(x+4i).
\label{d2frel}
\end{equation}

For $M>2$, such a simple polynomial expression is not
available and it reads explicitly,
\begin{eqnarray}
\sum_{k=0}^{M} \phi(x+2i k)&=&\frac{\pi}{2},   \label{dmrel} \np
\phi(x) &=&\arcsin \frac{1}{\sqrt{{\cal D}_M(x){\cal D}_M(x+2i)}}.
\label{phidef}  \np
\end{eqnarray}

On the other hand, transfer matrices are introduced in the analysis
of statistical mechanics\cite{Baxbook}, 
integrable structures in $c<1$ CFT \cite{BLZ1, BLZ2}
and so on \cite{FLS}.
We do not specify its precise definition.
(We refer to
\cite{Baxbook,KNS, BLZ1} for interested readers .)
For our purpose, the following facts are sufficient.
Let the deformation parameter $q$ be
${\rm e}^{i\pi \beta^2}$.
We denote  a  $U_q(\widehat{\mathfrak{sl}}_2)$  module $W_j(\lambda)$,
which corresponds 
to the $j+1$ dimensional module of $U_q(\mathfrak{sl}_2)$.
The associated ("Drinfel'd") polynomial is given by
\begin{equation}
P(\lambda') =(1-q^{j-1}\lambda \lambda')(1-q^{j-3}\lambda \lambda')
\cdots (1-q^{-j+1}\lambda \lambda').
\end{equation}
See \cite{ ChP} for precise definitions.
Taking trace of monodromy operator over $W_j(\lambda)$,
one can define  the transfer matrix ${\bf T}_j(\lambda)$.
${\bf T}_j(\lambda)$ constitutes a commutative family and satisfy
"$T-$ system",
\begin{equation}
{\bf T}_j(q \lambda) {\bf T}_j(q^{-1} \lambda)=
{\bf I}+{\bf T}_{j+1}(\lambda){\bf T}_{j-1}(\lambda),
\qquad j=1,2, \cdots, 
\label{tsys1}
\end{equation}
and ${\bf T}_0={\bf I}$.
(Note the suffix $j$ and the normalization
of $\lambda$ are defined differently from \cite{BLZ1}.) 

As we are considering these operators on their common eigenvector space,
we will use the same symbol ${\bf T}_j$ for its eigenvalue.

For $\beta^2 =\frac{1}{M+1}$, the above functional relations close finitely
due to the following property,
\begin{equation}
{\bf T}_{M-j}(\lambda) = {\bf T}_{M+j}(\lambda), \qquad j=1,\cdots, M
\label{dual}
\end{equation}
and ${\bf T}_{2M+1}(\lambda)=0$.

Again we adopt the "additive variable" $x$ rather than
"multiplicative  variable" $\lambda$,
 $T_j(x) = {\bf T}_j( e^{\pi x/(M+1)})$.
Then $T-$ system (\ref{tsys1}) reads
\begin{equation}
T_j(x+i) T_j(x-i)=1+T_{j+1}(x) T_{j-1}(x).
\label{tsys2}
\end{equation}
We also remark periodicity,
\begin{equation}
T_j(x+ (2M+2)i) = T_j(x).
\label{period}
\end{equation}
( The variable $\theta$ in \cite{DT} is related to $x$ by
$\theta =x \pi/2M$.)

In \cite{KP},\cite{KNS}, it has been shown 
that the substitution of $Y_j(x) = T_{j-1}(x) T_{j+1}(x)$ into 
(\ref{tsys2}) yields the well-known $Y-$ system \cite{AlZ}.
The solution to $Y-$ or $T-$ system is not necessarily unique.
One needs to know zeros or singularities of
$Y_j(x)$, or equivalently,$T_j(x)$ in 
a "physical strip" ($\Im x \in [-1, 1]$) to fix a solution.
With this knowledge, one reaches 
the Thermodynamic Bethe Ansatz (TBA) equation, which yields a unique solution.

Dorey and Tateo showed, for $M=2$, ${\cal D}_2(x)$ and $T_2(x)$ satisfy
the same functional relation (\ref{d2frel}).
The coincidence carries forward.
With some additional tuning of parameters, they share same
analytic structure, which validates 
${\cal D}_2 (x) =T_2(x)$.
For $M>2$ they did present numerical evidences to support a
conjecture ${\cal D}_M(x) =T_M(x)$ instead of
proving that they satisfy the same functional relation (\ref{dmrel}).

In the following we will supply the proof.
The idea is to utilize 
the short exact sequence of  $U_q(\widehat{\mathfrak{sl}_2})$
 in \cite{ChP}.
($T-$ system is one of the simplest consequences of it.)
The short exact sequence reads,
\begin{eqnarray}
0 & &\longrightarrow W_{\alpha-p}(\lambda q^{-p}) 
             \otimes W_{\beta-p}(\lambda' q^{-p})
  \longrightarrow W_{\alpha}(\lambda )\otimes W_{\beta}(\lambda' ) \np
 & &\longrightarrow W_{p-1}(\lambda q^{\alpha-p+1})\otimes
                      W_{\alpha+\beta-p+1}(\lambda' q^{-(\alpha-p+1)}) 
    \longrightarrow 0  \np                  
 & & \text{ for } \frac{\lambda'}{\lambda}=q^{\alpha+\beta-2p+2}.
\label{exactsq}
\end{eqnarray}
We abbreviate these modules to $W_0 \sim W_5$, and the corresponding
transfer matrix $T_{W_i}$ (trace of monodromy operator over $W_i$). 
Then the consequence of (\ref{exactsq}) is,
\begin{equation}
0=T_{W_0} T_{W_1} -T_{W_2} T_{W_3} +T_{W_4} T_{W_5}.
\end{equation}
In the additive variable, the equivalent "generalized
$T-$system" reads,
\begin{eqnarray}
T_{\alpha}(x) T_{\beta}(x+(\alpha+\beta-2p+2)i) & &
=T_{\alpha-p}(x-ip) T_{\beta-p}(x+i(\alpha+\beta-p+2)) \np
&+&
T_{p-1}(x+i(\alpha-p+1)) T_{\alpha+\beta-p+1}(x+i(\beta-p+1)).
\label{chariP}
\end{eqnarray}
One substitutes $\alpha=\beta=p=j$ to recover (\ref{tsys2}).
We refer to the above identity by $I(\alpha,\beta, p, x)$.

We first give the statement.
\begin{thm}
Let $\psi(x)$ be $\phi(x)$ in (\ref{phidef}) replacing
${\cal D}_M(x)$ by $T_M(x)$,
\begin{equation}
\psi(x) =\arcsin \frac{1}{\sqrt{T_M(x)T_M(x+2i)}}.
\end{equation}
Then we have,
\begin{equation}
\sum_{k=0}^{M} \psi(x+2i k)=\frac{\pi}{2}.
\label{psirel}
\end{equation}
\label{mainth}
\end{thm}

We prove the above theorem in an equivalent form, 
\begin{equation}
\begin{cases}
\cos(\psi(x)+\psi(x+2i)+\cdots+ \psi(x+(2M-4)i) ) = \\
 \qquad \sin(\psi(x+(2M-2)i)+\psi(x+2Mi))& \text { $M$ odd, }\\
\cos(\psi(x)+\psi(x+2i)+\cdots+ \psi(x+(2M-2)i) ) =  \\
 \qquad  \sin(\psi(x+2Mi)))& \text { $M$ even, }\\
\end{cases}
\label{trigrel}
\end{equation}
following \cite{V2}.
To be precise, the condition (\ref{trigrel}) literally
leaves multiples of $2\pi $ indeterminate in rhs of (\ref{psirel}).
This can be however fixed from the asymptotic value 
$T_M(|x| \rightarrow \infty ) = 1/\sin (\pi/(2M+2))$,
which can be derived from the algebraic relation 
(\ref{tsys2}) by sending $x\rightarrow \infty$.
We verify (\ref{trigrel})  coincides with  (\ref{psirel}).

To show (\ref{trigrel}),  we prepare few lemmas,
\begin{lem}
\begin{eqnarray}
\sin(\psi(x)+\psi(x+2i))&=&
\frac{T_1(x+i(M+3))}{\sqrt{T_M(x) T_M(x+4i)}} \np
\cos(\psi(x)+\psi(x+2i))&=&
\frac{T_{M-2}(x+2i)}{\sqrt{T_M(x) T_M(x+4i)}}.\np
\end{eqnarray}
\label{pair1}
\end{lem}
{\it Proof }  We first note 
\begin{eqnarray}
\cos(\psi(x)) &=&\sqrt{1-\sin^2 (\psi(x))}  
=\sqrt{1-\frac{1}{T_M(x)T_M(x+2i)} } \np
&=& \sqrt{\frac{T_{M-1}(x+i)T_{M+1}(x+i)}{ T_M(x)T_M(x+2i) }} 
= \frac{T_{M-1}(x+i)}{\sqrt{T_M(x)T_M(x+2i)}},
\end{eqnarray}
where (\ref{tsys2}) and (\ref{dual}) are used in the last two equalities.
By expanding the lhs of the first equation in Lemma \ref{pair1},
we have,
\begin{eqnarray}
\sin(\psi(x)+\psi(x+2i)) &=& \sin(\psi(x))\cos(\psi(x+2i)) +
                              \sin(\psi(x+2i))\cos(\psi(x))  \np
 &=& \frac{T_{M-1}(x+3i)+T_{M-1}(x+i)}{T_M(x+2i) \sqrt{T_M(x)T_M(x+4i)}}\np
 &=&  \frac{T_1(x+i(M+3))}{\sqrt{T_M(x) T_M(x+4i)}} \np
\end{eqnarray}
where we have applied $I(M,1,1,x+2i)$,
\begin{eqnarray}
T_M(x+2i)T_1(x+i(M+3)) &=& T_{M-1}(x+i)+T_{M+1}(x+3 i) \np
                       &=& T_{M-1}(x+i)+T_{M-1}(x+3 i).
\end{eqnarray}
Second relation is similarly proved.$\Box$

We further generalize the expansion of trigonometric
functions with even arguments more than 2.
\begin{lem}
Let $\ell$ be an odd integer.
We have the following relation.
\begin{eqnarray}
& &\left(
  \begin{array}{c}
           \cos (\psi(x)+\psi(x+2i)+\cdots +\psi(x+2\ell i)) \\
           \sin (\psi(x)+\psi(x+2i)+\cdots +\psi(x+2\ell i))      \\
  \end{array}
\right )  \np
& &=
\frac{1}{\sqrt{T_M(x)T_M(x+i(2\ell+2))}}
\prod_{k=1}^{(\ell-1)/2}
 \frac{ \mathfrak{L} (x+(4k-4)i)}{ T_M (x+4k i)}
\left(
  \begin{array}{c}
           T_{M-2}(x+2\ell i) \\
           T_1(x+i(M+2\ell+1)    \\
  \end{array}
\right )        \np
&&     \mathfrak{L}(x) :=
          \left(
          \begin{array}{cc}
           T_{M-2} (x+2i),&   -T_1(x+i(M+3) )\\
           T_1(x+i(M+3)) ,&    T_{M-2} (x+2i) \\
          \end{array}
            \right )
            \label{allrec}
\end{eqnarray}
where the order of the operator product should be understood as,
$$
\mathfrak{L}(x) \mathfrak{L}(x+4i) \cdots \mathfrak{L}(x+(2\ell-6)i).
$$
\end{lem}
{\it Proof }: This is easily shown by iterative applications
of the recursion relation,
\begin{eqnarray}
& &\left(
  \begin{array}{c}
           \cos (\psi(x)+\psi(x+2i)+\cdots +\psi(x+2\ell i)) \\
           \sin (\psi(x)+\psi(x+2i)+\cdots +\psi(x+2\ell i))      \\
  \end{array}
\right )  \np
& &=
\frac{1}{\sqrt{T_M(x)T_M(x+4i)}}
\mathfrak{L}(x)
\left(
  \begin{array}{c}
             \cos (\psi(x+4i)+\psi(x+6i)+\cdots +\psi(x+2\ell i))  \\
            \sin (\psi(x+4i)+\psi(x+6i)+\cdots +\psi(x+2\ell i))     \\
  \end{array}
\right ) ,
\end{eqnarray}
which follows from Lemma (\ref{pair1}). $\Box$

The above recursion procedure is regarded as the forward propagation.
Next let us perform the back-propagation procedure: we apply
matrices $\mathfrak{L}$ on the column vector.
We observe a simple pattern there, which can be summarized as
the following lemma.

\begin{lem}
We introduce a vector $v_t$ by
\begin{equation}
v_t :=\left(
  \begin{array}{c}
            T_{M-2-2t}(x-(6+2t)i)  \\
            T_{2t+1}(x+(M-5-2t)i)  \\
  \end{array}
\right ) .
\label{defvt}
\end{equation}
Then the following  back-propagation relation holds,
\begin{equation}
\mathfrak{L}(x+i(2M-10-4t)) v_t =
T_{M}(x-(4t+8)i) v_{t+1}.
\label{vtvt1}
\end{equation}
\end{lem}
{\it Proof }: The first component of lhs in (\ref{vtvt1}) reads
\begin{eqnarray}
& &T_{M-2}(x-(10+4t)i) T_{M-2-2t}(x-(6+2t)i)  \np
 & &- T_1(x+i(M-9-4t)) T_{2t+1}(x-i(M+2t+7)),
\label{firstc}
\end{eqnarray}
where we have applied the periodicity (\ref{period}) to the last component.
By use of $I(M-2, M-2t-2, M-2t-3,x-(4t+10)i)$, one finds (\ref{firstc})
equals to $T_{M}(x-(4t+8)i) T_{M-2-2(t+1)}(x-i(2(t+1)+6)i)$, which 
is nothing but the first component  of rhs.
Similarly one applies $I(M,2t+3,2,x-(4t+8)i)$ to the second component
of rhs in (\ref{vtvt1}), leading to the equality.$\Box$

We shall fix the relation between $\ell$ and $M$ as follows,
\begin{equation}
\ell = \begin{cases}
          M-2,& \text{if $M=$ odd } \\
          M-1,&  \text{ if $M=$ even }
       \end{cases}
\end{equation}

Then our final lemma is
\begin{lem}
Under the above relation between $\ell$ and $M$, one has
\begin{equation}
\cos(\psi(x)+\cdots \psi(x+2\ell i))=
\begin{cases}
          \frac{T_1(x-i(M+3))}{\sqrt{T_M(x) T_M(x+(2M-2)i)}},&
            \text{ $M$ odd } \\
         \frac{1}{\sqrt{T_M(x) T_M(x+2Mi)}},&
            \text{ $M$ even. }
\end{cases}
\end{equation}
\label{flemma}
\end{lem}
{\it Proof }: Let us apply $\mathfrak{L}$'s  to the vector in (\ref{allrec}).
For $M$ odd, the initial vector reads,
\begin{equation}
\left (
 \begin{array}{c}
           T_{M-2}(x-6i) \\
           T_1(x+i(M-5)i)   \\
  \end{array}
  \right )
\end{equation}
which is nothing but $v_{t=0}$ in (\ref{defvt}).
The product of matrices in (\ref{allrec}) is 
\begin{equation}
 \mathfrak{L}(x) \mathfrak{L}(x+4i)  \cdots \mathfrak{L}(x+(2M-14)i)   
 \mathfrak{L}(x+(2M-10)i).
\label{Lpmodd}  
\end{equation}
Thus one can apply (\ref{vtvt1}) recursively to find
\begin{equation}
\mathfrak{L}(x)  \cdots \mathfrak{L}(x+(2M-10)i) v_0  
= \prod_{j=1}^{(M-3)/2} T_M(x-(4+4j)i) 
\left (
\begin{array}{c}
           T_{1}(x-i(M+3)) \\
           T_{M-2}(x-2i)   \\
  \end{array}
\right )
\label{prodModd}
\end{equation}
Substituting (\ref{prodModd}) into (\ref{allrec}), and after
rearrangement using (\ref{period}) one arrives at the odd case
of Lemma \ref{flemma} from the first component. 
For $M$ even, initial vector is 
$v'_{t=0}=v_{t=0} (x\rightarrow x+2i)$.
Similarly, the product of $\mathfrak{L}$ is given by
$x\rightarrow x+2i$ in (\ref{Lpmodd}).
The result of the application reads
\begin{equation}
 \cdots \mathfrak{L}(x+(2M-12)i) \mathfrak{L}(x+(2M-8)i) v'_0  
= \prod_{j=1}^{(M-2)/2} T_M(x-(2+4j)i) 
\left (
\begin{array}{c}
           T_{0}(x-(M+2)i) \\
           T_{M-1}(x-i)   \\
  \end{array}
\right )
\label{prodMeven}
\end{equation}
Again the  substitution of  (\ref{prodMeven}) into (\ref{allrec})
leads to  Lemma \ref{flemma} for $M$ even case. $\Box$

\noindent{\it Proof of Theorem \ref{mainth}. }: 
Now the left hand side of (\ref{trigrel}) is explicitly written in terms of
$T-$functions in  Lemma \ref{flemma}. It remains to check that it coincides with
 rhs. This can be easily done by (\ref{pair1}) or from the definition of
 $\psi(x)$ itself. $\Box$

As is noted previously,
the  common functional relation does not grantee the equality,
$T_M(x)={\cal D}_M(x)$: one needs further knowledge on their analytic 
structures. 
In this respect, we shall entirely depend on the argument
in \cite{DT}.  In the TBA equation originated from $T-$ system,
one shall take the massless drive terms,
$m_a r {\rm e}^{\pi x/2M}, (a=1, \cdots 2M-1)$ and 
setting $m_M r =
\pi^{1/2} \Gamma(\frac{1}{2M})/(M \Gamma(\frac{3}{2}+\frac{1}{2M}))$.
Then ${\cal D}_M(x)$ and $T_M(x)$ shares same analytical properties:
Both of them have same "asymptotic value "  and  have 
zeros on $\Im x= \pm(M+1)$. 
The latter is consistent with a property of 
the Schr{\"o}dinger operator that eigenstates are all bounded
so $E_k <0$.
Thus one concludes  the equality, ${\cal D}_M(x)=T_M(x)$
 
Summarizing, we have proven one of conjectures in \cite{DT} that 
$T_M(x)$ actually shares same functional relation with ${\cal D}_M(x)$.
The proof utilizes the exact sequence of $U_q(\widehat{\mathfrak{sl}}_2)$.
This makes us expect further deep connections between the
anharmonic oscillator and quantum integrable structures.

\section*{Acknowledgments}
The author thanks A. Kuniba for calling his attention to \cite{DT}
and for useful comments.
He also thanks V V Bazhanov, T Miwa, Y Pugai and Z Tsuboi for
comments.

\end{document}